\documentclass[a4paper,11pt]{article}
\pdfoutput=1 
\usepackage{jcappub} 
\usepackage[T1]{fontenc} 

\title{\boldmath Muon g-2 and galactic centre $\gamma$-ray excess in a scalar extension of the 2HDM type-X}

\usepackage[utf8]{inputenc}
\usepackage[english]{babel}
\usepackage{graphicx} 
\usepackage{placeins}
\usepackage{subcaption}
\usepackage{mathrsfs}
\usepackage{xcolor}

\newcommand{\hc}[1]{#1^{\dagger}}
\newcommand{\abs}[1]{|#1|}
\newcommand{\tb}{t_{\beta}}
\newcommand{\dmu}{\Delta a_\mu}

\newcommand{\td}{\mathrm{d}}
\newcommand{\sv}{\left\langle\sigma v\right\rangle}

\author[a,1]{Andi Hektor,\note{Corresponding author.}}
\author[a]{Kristjan Kannike,}  
\author[a,b]{and Luca Marzola}

\affiliation[a]{National Institute of Chemical Physics and Biophysics; R\"{a}vala pst. 10,\\ 10143 Tallinn, Estonia}
\affiliation[b]{Institute of Physics, University of Tartu; Ravila 14c, 50411 Tartu, Estonia
}

\emailAdd{andi.hektor@cern.ch}
\emailAdd{kristjan.kannike@cern.ch}
\emailAdd{luca.marzola@ut.ee}

\abstract{
We consider an extension of the lepto-specific 2HDM with an extra singlet $S$ as a dark matter candidate. Taking into account theoretical and experimental constraints, we investigate the possibility to address both the $\gamma$-ray excess detected at the Galactic centre and the discrepancy between the Standard Model prediction and experimental results of the anomalous magnetic moment of the muon. Our analyses reveal that the $SS \to \tau^+ \tau^-$ and $SS \to b \bar b$ channels reproduce the Galactic centre excess, with an emerging dark matter candidate which complies with the bounds from direct detection experiments, measurements of the Higgs boson invisible decay width and observations of the dark matter relic abundance. Addressing the anomalous magnetic moment of the muon imposes further strong constraints on the model. Remarkably, under these conditions, the $SS \to b \bar b$ channel still allows for the fitting of the Galactic centre. We also comment on a scenario allowed by the model where the $SS \to \tau^+ \tau^-$ and $SS \to b \bar b$ channels have comparable branching ratios, which possibly yield an improved fitting of the Galactic centre excess.
}
\begin{document}
\maketitle 
\flushbottom

\section{Introduction}
\label{sec:intro}


The matter content of the Universe is dominated by a weakly interacting component, the Dark Matter (DM), which accounts for about 26~\% of the energy density of the former at the present time. According to the current paradigm, DM consists of a relic abundance of weakly-interacting massive particles (WIMPs). This scheme is motivated by the fact that particles with masses and annihilation cross sections set by the electroweak scale, a known scale of Nature, yield through the freeze-out mechanism\footnote{For a review see~\cite{Jungman:1995df,Bertone:2004pz}.} a relic abundance comparable with the observed DM one. Given that the freeze-out mechanism is a natural consequence of an expanding Universe and that new weakly-interacting particles arise in well motivated extensions of the Standard Model (SM), e.g. super-symmetric theories, the above observation is referred to in literature as the  ``WIMP miracle''. The WIMP paradigm motivates the experimental efforts behind different DM searches: for instance, the direct detection experiments rely on the possible elastic scattering between DM and Standard Model (SM) particles mediated by the weak force. Likewise, the weak interactions could allow for the DM production at colliders, motivating the dedicated collider searches. Finally, the annihilation of DM particles in space, mediated by the same force, provide the basis for DM indirect detection. Regarding the latter possibility, notice that DM annihilations take place only in regions where the DM abundance is sizeable and, therefore, possible annihilation signals are expected from dense DM regions as the Galactic Centre (GC) and the dwarf satellite galaxies of the Galaxy. In addition to that, DM annihilations in space yield large scale effects that can be investigated, for example, by their impact on the Cosmic Microwave Background radiation. The latter is especially constricting for light WIMPs, $\lesssim10$~GeV~\cite{Galli:2009zc, Slatyer:2009yq, Huetsi:2009ex, Cirelli:2009bb, Hutsi:2011vx, Evoli:2012qh, Madhavacheril:2013cna, Ade:2015xua}.


In this light, the DM halo at GC should then provide the strongest indirect detection signals and, interestingly, in 2009 a spatially extended $\gamma$-ray excess was detected in this region by the Fermi LAT telescope in the energy range of 1-5~GeV~\cite{Atwood:2009ez}. We refer to this signal, confirmed by many other studies~\cite{Goodenough:2009gk, Hooper:2010mq, Abazajian:2010zy, Boyarsky:2010dr, Hooper:2011ti, Abazajian:2012pn, Gordon:2013vta, Macias:2013vya, Abazajian:2014fta, Daylan:2014rsa, Lacroix:2014eea, Calore:2014nla}, as the Galactic Centre Excess (GCE). Clearly the GC harbours an extremely rich environment populated with stars, stellar relics, cosmic rays, dust, gas and the central black hole and, so far, it has not been possible to clearly disentangle the potential annihilation signal from the partially known astrophysical background due, for instance, to millisecond pulsars~\cite{Yuan:2014rca} and ultra-energetic events form the past~\cite{Petrovic:2014uda}. Nevertheless, the DM annihilation interpretation is supported by the estimates of the required annihilation cross section, matching the order of the thermal freeze-out one, as well as by the morphology of the signal region. 

When interpreting the detected GCE in the light of the WIMP paradigm it is necessary to address the strong constraints from the direct detection experiments XENON100~\cite{Aprile:2012nq} and LUX~\cite{Akerib:2013tjd}, which disfavour the characteristic weak-scale values of scattering cross section. This difficulty has pushed the scientific community to consider alternative models that possess an annihilation cross section large enough to explain the detected signal, but present a suppressed DM-nucleon scattering cross section. An example is provided by the so-called Coy Dark Matter \cite{Boehm:2014hva} models, in which the DM particle is a Dirac fermion $\chi$ interacting with the SM particles through a light ($\sim$10~GeV) pseudoscalar mediator. By assuming that the couplings to the SM particles are proportional to the corresponding Higgs Yukawa couplings, as motivated by minimal flavour violation \cite{DAmbrosio:2002ex}, the model avoids the tight direct detection constraints\footnote{Interestingly, a light mediator can be exploited for the annual modulation signal of DAMA~\cite{Arina:2014yna}.}. Furthermore, if the DM particles are lighter than the top quark, the dominant annihilation channel is $\chi\chi \to b \bar b$ and with the choice $m_\chi \in [40,55]$~GeV, the GCE can be fitted for a natural value of the DM annihilation cross section~\cite{Gordon:2013vta, Macias:2013vya, Abazajian:2014fta, Daylan:2014rsa, Lacroix:2014eea, Calore:2014nla}.

Alternatively, it is possible to fit the signal with the $\chi\chi \to \bar{\tau}\tau$ channel by adopting a DM mass $m_\chi \simeq 10$~GeV. In this case, assuming leptophilic SM-pseudoscalar couplings proportional to the corresponding lepton Yukawa couplings also removes the tension between the SM prediction and the measurement of anomalous magnetic moment of the muon  $a_\mu$~\cite{Hektor:2014kga}. In this regard, the long-standing $\sim 3\sigma$ discrepancy\footnote{For a review see~\cite{Beringer:1900zz} and references therein. We consider a conservative value for the statistical significance of the mentioned discrepancy  \cite{Benayoun:2015gxa}.} that afflicts this quantity can be regarded, along with the origins of DM, neutrino masses and baryon asymmetry of the Universe, as a compelling experimental evidence that points to physics beyond the SM. In this paper we adopt the above point of view and fit the GCE within the framework of the extended Two Higgs Doublet Model (2HDM), investigating also the consequences of the implied particle content on the anomalous magnetic moment of the muon. 


The 2HDM \cite{Gunion:1989we,Branco:2011iw} is a well known extension of the SM in which a second scalar doublet field is added to the particle content of the theory. The latter transforms under the symmetries of the SM in the same way as the original Higgs doublet and, depending on the model, is allowed to develop a non-zero vacuum expectation value (VEV). The SM-like Higgs boson, discovered at the LHC \cite{Chatrchyan:2012xdj,Aad:2012tfa}, is a combination of the two doublets. Barring models that induce flavour-changing neutral currents either at the tree or loop level, there are four possible ways to assign the Yukawa couplings with the SM fermions to the two scalar doublets \cite{Aoki:2009ha,Gunion:2002zf,Branco:2011iw}. As shown in table~\ref{tab:2HDM:types}, these correspond to  different $\mathbb{Z}_{2}$ parity assignments, which mark the four possible 2HDM variants: type-I, II, X (or ``lepto-specific'') and Y (or ``flipped''). The model in which the additional Higgs doublet does not acquire a VEV nor couples to the fermions is instead called the inert doublet model~\cite{Barbieri:2006dq,Ma:2006km}.

\begin{table}[htp]
\caption{$\mathbb{Z}_{2}$ parity and Yukawa enhancement factors $y_{x}^{i}$.}
\begin{center}
\begin{tabular}{|cccccccccc|}
\hline
& $H_{1}$ & $H_{2}$ & $u_{R}^{c}$ & $d_{R}^{c}$ & $\ell_{R}^{c}$ & $Q_{L}, L_{L}$ & $y_{u}$ & $y_{d}$ & $y_{\ell}$  \\
\hline
Type-I & $+$ & $-$ & $-$ & $-$ & $-$ & $+$ & $\cot \beta$ & $\phantom{-}\cot \beta$ & $\phantom{-}\cot \beta$ \\
Type-II & $+$ & $-$ & $-$ & $+$ & $+$ & $+$ & $\cot \beta$ & $-\tan \beta$ & $-\tan \beta$ \\
Type-X & $+$ & $-$ & $-$ & $-$ & $+$ & $+$ & $\cot \beta$ & $\phantom{-}\cot \beta$ & $-\tan \beta$ \\
Type-Y & $+$ & $-$ & $-$ & $+$ & $-$ & $+$ & $\cot \beta$ & $-\tan \beta$ & $-\tan \beta$ \\
\hline
\end{tabular}
\end{center}
\label{tab:2HDM:types}
\end{table}%

In this paper we focus on an extension of the lepto-specific 2HDM, where a further real singlet scalar field $S$ is added to play the role of dark matter~\cite{Silveira:1985rk,McDonald:1993ex,Burgess:2000yq,Barger:2007im,Barger:2008jx,Gonderinger:2009jp,Cai:2011kb,Chen:2012faa,Gonderinger:2012rd} (see~\cite{Cline:2013gha} for a recent review). Whereas such a construction has been previously considered in literature~\cite{Boucenna:2011hy,Okada:2013bna,Bonilla:2014xba} (or as an effective field theory~\cite{Alves:2014yha}), the novelty of our work is in the detailed analysis of the GCE fit, which accounts for the latest bounds that the mentioned direct detection experiments yield. We furthermore scrutinise the impact of the extended particle content presented by the model on the anomalous magnetic moments of the electron and muon, investigating whether the same values of the parameters that fit the GCE  reduce, as well, the tension that afflict the SM prediction of the latter.

The paper is organised as follows: in the next section we provide the details of the considered model, in order to address in section~\ref{sec:Perturbative Unitarity and bounds from precision Electroweak measurements} the bounds that perturbativity, perturbative unitarity, vacuum stability and Electroweak precision data impose. Section~\ref{sec:gamma_ray} is dedicated to fitting the GCE, while in section~\ref{sec:gm2} we discuss the muon anomalous magnetic moment in the adopted framework. We summarise our results in section~\ref{sec:Conclusions}.

\section{The considered model} 
\label{sec:The considered model}

The Higgs sector of the 2HDM presents two $SU(2)$ scalar doublets $H_{1}$ and $H_{2}$. In order to avoid flavour-changing neutral currents, it is customary to impose a $\mathbb{Z}_{2}$ symmetry on the scalar potential. As said before, there are four possible ways to assign a $\mathbb{Z}_{2}$ parity to the fields, summarised in table~\ref{tab:2HDM:types}.

In the lepto-specific, or type-X 2HDM, $H_{1} \equiv H_{L}$ couples only to leptons while $H_{2} \equiv H_{Q}$ only to quarks. Our scalar sector contains, in addition to the two doublets, a real singlet $S$ that is our dark matter candidate.\footnote{Similar models have been explored in the context of DM stabilised by $\mathbb{Z}_{N}$ symmetries in \cite{Belanger:2014bga,KANNIKE:2014mka}.} We forbid explicit CP-violation but allow for the soft breaking of the $\mathbb{Z}_{2}$ symmetry that characterises the considered model. The most general scalar potential in line with these assumptions is then
\begin{equation}
\begin{split}
  V &= -\mu_{L}^{2} \hc{H_{L}} H_{L} - \mu_{Q}^{2} \hc{H_{Q}} H_{Q} - \mu_{LQ}^{2} (\hc{H_{L}} H_{Q} + \hc{H_{Q}} H_{L}) + \frac{1}{2} \mu_{S}^{2} S^{2}
  + \lambda_{S} S^{4} \\
  &+ \lambda_{1} (\hc{H_{L}} H_{L})^{2} + \lambda_{2} (\hc{H_{Q}} H_{Q})^{2} 
  + \lambda_{3} (\hc{H_{L}} H_{L}) (\hc{H_{Q}} H_{Q}) + \lambda_{4} (\hc{H_{L}} H_{Q}) (\hc{H_{Q}} H_{L}) \\
  &+ \frac{1}{2} \lambda_{5} \left[ (\hc{H_{Q}} H_{L})^{2} + (\hc{H_{L}} H_{Q})^{2} \right] 
  + \lambda_{SL} \hc{H_{L}} H_{L} S^{2} + \lambda_{SQ} \hc{H_{Q}} H_{Q} S^{2},
\end{split}
\label{eq:potential}
\end{equation}
where both $\mu_{LQ}^{2}$ and $\lambda_{5}$ are real and the masses $\mu_{LQ}^{2}$ are responsible for the soft breaking of the $\mathbb{Z}_{2}$ symmetry.

As usual we parametrise the degrees of freedom contained in the Higgs doublets as\footnote{We follow the conventions of \cite{Abe:2015oca}.}

\begin{equation}
	H_i = 
	\begin{pmatrix}
		h_i^+ \\ (v_i + h_i -ia_i)/\sqrt2
	\end{pmatrix},
\end{equation}
where $v_{1}$ and $v_{2}$ are the VEVs of the two scalar doublets. The particle content of the SM is then effectively extended to accommodate an extra neutral scalar $H$, the charged scalars $H^\pm$ and a neutral pseudoscalar $A$. We indicate with $h$ the usual SM Higgs boson.

In the following, we choose as free parameters the SM-like Higgs mass $m_{h}= 125.09~\text{GeV}$ \cite{Aad:2015zhl}, the pseudoscalar mass $m_{A}$, the neutral scalar mass $m_{H}$, the charged Higgs mass $m_{H^{\pm}}$, the DM mass $m_{S}$, $\lambda_{1}$, $\lambda_{2}$, $\lambda \equiv \lambda_{3} + \lambda_{4} + \lambda_{5}$, $\lambda_{S}$, $\lambda_{SL}$ and $\lambda_{SQ}$, the neutral scalar mixing angle $\alpha$, together with $v^{2} = v_{1}^{2} + v^{2} = 246.2$~GeV and $\tb \equiv \tan \beta = v_{2}/v_{1}$.

We can express
\begin{align}
  \lambda_{1} v^{2} &= (m_{H}^{2} \tb^{2} + m_{h}^{2}) s_{\beta - \alpha}^{2} 
  + (m_{H}^{2} + m_{h}^{2} \tb^{2}) c_{\beta - \alpha}^{2} 
  + 2 (m_{H}^{2} - m_{h}^{2}) \tb s_{\beta - \alpha}^{2} c_{\beta - \alpha}^{2} 
  - M^{2} \tb^{2},\label{eq:lambda:1:param}
  \\
  \lambda_{2} v^{2} &= \left( \frac{m_{H}^{2}}{\tb^{2}} + m_{h}^{2} \right) s_{\beta - \alpha}^{2} 
  + \left( m_{H}^{2} + \frac{m_{h}^{2}}{\tb^{2}} \right) c_{\beta - \alpha}^{2} 
  + 2 (m_{H}^{2} - m_{h}^{2}) \tb s_{\beta - \alpha}^{2} c_{\beta - \alpha}^{2} 
  - \frac{M^{2}}{\tb^{2}},
  \\
  \lambda v^{2} &= 2 (m_{h}^{2} - m_{H}^{2}) s_{\beta - \alpha} + 2 (m_{H}^{2} - m_{h}^{2}) c_{\beta - \alpha}
  + 2 \left( \frac{1}{\tb} - \tb \right) (m_{h}^{2} - m_{H}^{2}) s_{\beta - \alpha} c_{\beta - \alpha}
  \\
  & + 2 M^{2}, \notag
  \\
  \lambda_{3} &= \lambda - \lambda_{4} - \lambda_{5},
  \\
  \lambda_{4} v^{2} &= M^{2} + m_{A}^{2} - 2 m_{H^{\pm}}^{2},
  \\
  \lambda_{5} v^{2} & = M^{2} - m_{A}^{2},
\end{align}
where $M^{2} \equiv \mu_{LQ}^{2} \left(\tb + \frac{1}{\tb} \right)$,  $s_{\beta - \alpha} \equiv \sin (\beta - \alpha)$ and $c_{\beta - \alpha} \equiv \cos (\beta - \alpha)$ \cite{Abe:2015oca}. Because the coupling $\lambda_{1}$, as given by eq.~\eqref{eq:lambda:1:param}, tends to large for increasing $t_{\beta}$, we adopt $\lambda_{1}$ as an input parameter instead of $\mu_{LQ}^{2}$, in virtue of eq.~\eqref{eq:lambda:1:param}.


\section{Theoretical and experimental constraints} 
\label{sec:Perturbative Unitarity and bounds from precision Electroweak measurements}

\subsection{Perturbativity and perturbative unitarity}

To satisfy a perturbativity constraint, we require the absolute values of all quartic couplings to be smaller than $4 \pi$.

We also consider perturbative unitarity. Because at high energy the scattering rates are dominated by quartic contact interactions, we require the eigenvalues of the $s$-wave $S$-matrix of two-to-two scalar boson states to satisfy 
\begin{equation}
  |a_{ij}| < \frac{1}{2}.
\end{equation}

For the 2HDM the unitarity constraints have been given in \cite{Kanemura:1993hm,Akeroyd:2000wc,Ginzburg:2003fe,Ginzburg:2005dt}. As in \cite{Belanger:2014bga}, we extend the method of \cite{Ginzburg:2003fe,Ginzburg:2005dt} to our field content. The eigenvalues are given by
\begin{align}
  16 \pi \Lambda_{21\pm}^{\text{even}} &= \lambda_{1} + \lambda_{2} \pm \sqrt{(\lambda_{1} - \lambda_{2})^{2} + \lambda_{5}^{2}}, \\
  16 \pi \Lambda_{21}^{\text{odd}} &= \lambda_{3} + \lambda_{4}, \\
  16 \pi \Lambda_{20}^{\text{odd}} &= \lambda_{3} - \lambda_{4}, \\
  16 \pi \Lambda_{01\pm}^{\text{even}} &= \lambda_{1} + \lambda_{2} \pm \sqrt{(\lambda_{1} - \lambda_{2})^{2} + \lambda_{4}^{2}}, \\
  16 \pi \Lambda_{00\pm}^{\text{odd}} &= \lambda_{3} + 2 \lambda_{4} \pm 2 \lambda_{5},
\end{align}
and the $\Lambda_{00\, 1,2,3}^{\text{even}}$ are given by $\frac{1}{16 \pi} \times$ the three roots of the polynomial equation
\begin{equation}
\begin{split}
  &2 x^3 - x^2 (12 \lambda_{1} + 12 \lambda_{2} + \sqrt{2} \lambda_{S}) + 
 12 \lambda_{2} \lambda_{SL}^2 + (2 \lambda_{3} + \lambda_{4}) (\sqrt{2} (2 \lambda_{3} + \lambda_{4}) \lambda_{S} - 
    4 \lambda_{SL} \lambda_{SQ}) \\
    &+ 
 12 \lambda_{1} (-3 \sqrt{2} \lambda_{2} \lambda_{S} + \lambda_{SQ}^2) - 
 2 x ((2 \lambda_{3} + \lambda_{4})^2 - 
    3 (12 \lambda_{1} \lambda_{2} + \sqrt{2} \lambda_{1} \lambda_{S} + 
       \sqrt{2} \lambda_{2} \lambda_{S}) \\
 &+ \lambda_{SL}^2 + \lambda_{SQ}^2) = 0.
\end{split}
\end{equation}

\subsection{Vacuum stability}

The potential is bounded below in the limit of large field values if the quartic couplings satisfy the copositivity conditions \cite{Kannike:2012pe}
\begin{align}
  \lambda_{1} &> 0, & \lambda_{2} &> 0, & \lambda_{S} &> 0, \\ 
  2 \sqrt{\lambda_{1} \lambda_{2}} + \lambda_{3} + \lambda_{4} - |\lambda_{5}| &> 0 &
  2 \sqrt{\lambda_{1} \lambda_{S}} + \lambda_{SL} &> 0 &
  2 \sqrt{\lambda_{2} \lambda_{S}} + \lambda_{SQ} &> 0 \\
  \span \span \span \span \span \span
  2 \sqrt{\lambda_{2} \lambda_{2} \lambda_{S}} 
  + \sqrt{\lambda_{1}} \lambda_{SQ} + \sqrt{\lambda_{2}} \lambda_{SL}
  + \sqrt{\lambda_{S}} (\lambda_{3} + \lambda_{4} - |\lambda_{5}|) & \\
  \omit  \span \omit \span \span \span \span
  + \sqrt{(2 \sqrt{\lambda_{1} \lambda_{S}} + \lambda_{SL}) (2 \sqrt{\lambda_{2} \lambda_{S}} + \lambda_{SQ})
  (2 \sqrt{\lambda_{1} \lambda_{2}} + \lambda_{3} + \lambda_{4} - |\lambda_{5}|)} &> 0. \notag
\end{align}

\begin{figure}[h]
\centering
\includegraphics[width=0.65\textwidth]{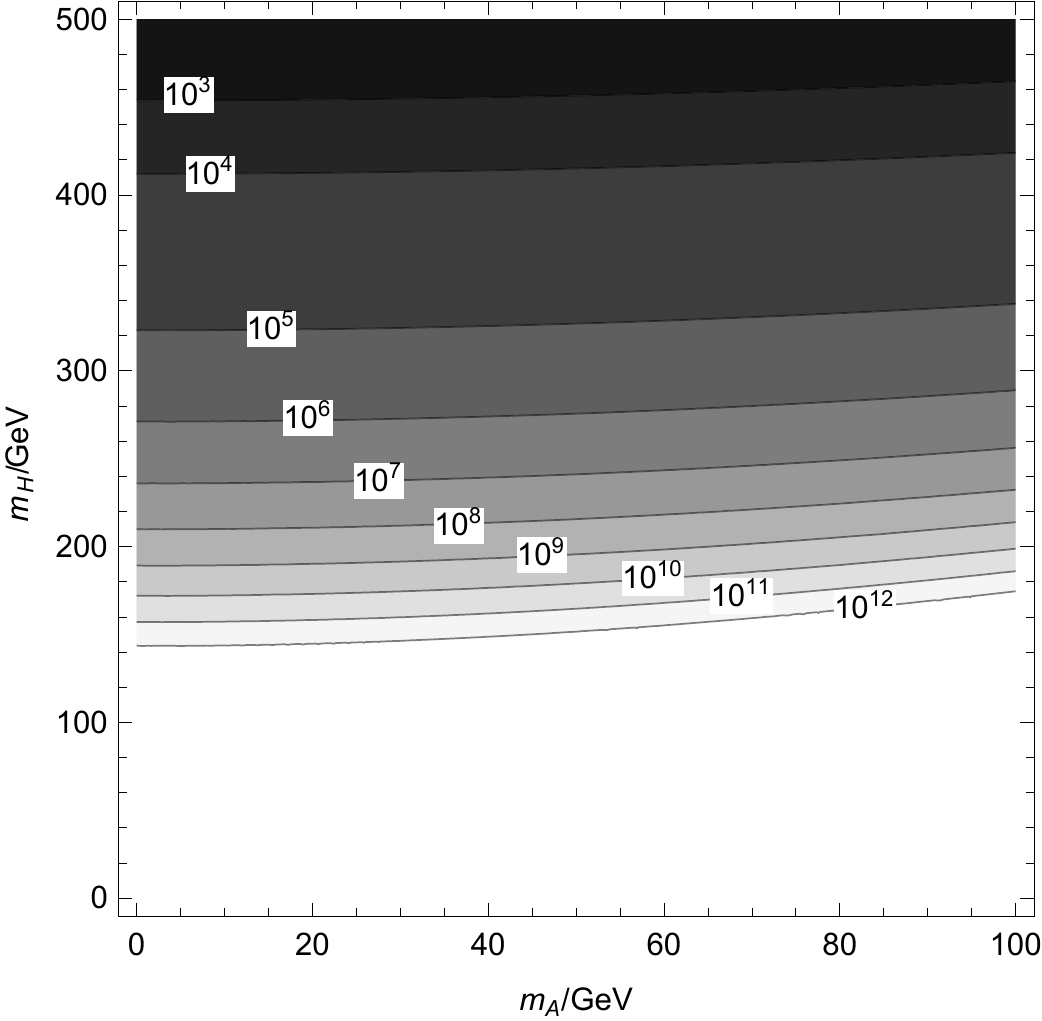}
\caption{The scale $\Lambda$ at which the model becomes non-perturbative as a function of $m_{H^{\pm}}= m_{H}$ vs. $m_{A}$.}\label{fig:constraints1}
\end{figure}

We have derived the renormalisation group equations (RGEs) for our model with the help of the PyR@TE package \cite{Lyonnet:2013dna}. The $\beta$-functions are given in the Appendix~\ref{app:RGEs}. We use the RGEs to find the scale $\Lambda$ at which the model becomes non-perturbative or loses vacuum stability. The results are shown in figure~\ref{fig:constraints1} as a function of $m_{A}$ and $m_{H}$.

\subsection{Electroweak precision data}

The electroweak oblique parameters $S$, $T$ and $U$ parametrise the effect of new physics on the gauge bosons two point functions \cite{Peskin:1991sw}.

The latest results from the GFitter group \cite{Baak:2014ora} are
\begin{equation}
  \Delta S = 0.05 \pm 0.11, \quad \Delta T = 0.09 \pm 0.13. 
\end{equation}

In the SM-like limit $\sin (\beta - \alpha) \to 1$, the electroweak precision $S$, $T$ and $U$ parameters \cite{Peskin:1991sw} reduce to those of the inert doublet model \cite{Barbieri:2006dq}. If $m_{A} \ll m_{Z} \ll m_{H^{\pm}} \approx m_{H}$, they are~\cite{Abe:2015oca}
\begin{equation}
  \Delta S \approx 0.022, \quad \Delta T \approx m_{H} \frac{\abs{m_{H^{\pm}} - m_{H}}}{32 \pi^{2} \alpha_{\text{em}} v^{2}}.
\end{equation}

Since the $T$ parameter constrains $\abs{m_{H} - m_{H^{\pm}}} = \mathcal{O}(10)$~GeV, in the following we always consider $m_{H^{\pm}} = m_{H}$ to comply with this constraint.

\subsection{$\lambda_{hA A}$ coupling}

The SM-like Higgs boson can decay into two pseudoscalars $A$ if $m_{A} < m_{h}/2$. Dedicated collider experiments bound the corresponding branching ratio to be smaller than $0.2$ \cite{Curtin:2013fra}, therefore we opt to set the coupling $\lambda_{hAA}$ to zero \cite{Abe:2015oca} by imposing that
\begin{equation}
	\label{eq:b-a_condition}
 \tan (\beta - \alpha) = \frac{M^{2} - m_{h}^{2}}{2 M^{2} - 2 m_{A}^{2} - m_{h}^{2}} \left(\tb - \frac{1}{\tb} \right).
\end{equation}

With the above prescription, the model still recovers the SM-like behaviour of $h$ in the limit $\tb\gg1$, allowing however for testable consequences on sign of the lepton Yukawa couplings \cite{Abe:2015oca}.  

\section{Fitting the $\gamma$-ray excess at the Galactic Centre} 
\label{sec:gamma_ray}

\subsection{The $\gamma$-ray excess at the Galactic Centre}

The data provided by the Fermi LAT $\gamma$-ray telescope~\cite{Atwood:2009ez} show a spatially extended $\gamma$-ray excess in the range 1-5~GeV at the GC~\cite{Goodenough:2009gk, Hooper:2010mq, Abazajian:2010zy, Boyarsky:2010dr, Hooper:2011ti, Abazajian:2012pn, Gordon:2013vta, Macias:2013vya, Abazajian:2014fta, Daylan:2014rsa, Lacroix:2014eea, Calore:2014nla}. In this paper we assume that the signal originates in DM annihilation, neglecting possible competing astrophysical processes as millisecond pulsars~\cite{Yuan:2014rca} and past energetic events in the GC~\cite{Petrovic:2014uda}. If the GCE is fitted by using a single channel, i.e. a single SM final state, the best fit is provided by $b\bar b$ and $c\bar c$.  Lighter or heavier quarks, vector and Higgs bosons as well as the $\tau^+\tau^-$ final state, nevertheless, are also viable options~\cite{Calore:2014nla}. We remark that if antiprotons are produced in the hadronization of final state particles~\cite{Cirelli:2010xx}, as in the case of quark, gluon, vector and Higgs boson channels, the GCE signal should be accompanied by an antiproton signal in cosmic rays. As the latter is not detected, the ``missing'' antiproton signal appears to be in tension with the DM interpretation of the GCE~\cite{Bringmann:2014lpa, Cirelli:2014lwa}. Nevertheless, this signal indeed depends strongly on the Galactic diffusion model, on which no consensus has yet been reached in literature~\cite{Evoli:2011id, Giesen:2015ufa, Evoli:2015vaa}. 
The missing antiproton problem can in principle be alleviated by invoking leptonic final states as $\tau^+\tau^-$. These, however, result in energetic electrons and positrons and therefore produce radio synchrotron radiation in the Galactic magnetic field. If the Galactic DM density profile presents a cusped core, as with the NFW profile, and the magnetic field in the interested region is sufficiently strong, $\sim$10~$\mu$G, the emitted synchrotron signal should dominate over the astrophysical background~\cite{Bertone:2008xr, Crocker:2010gy, Bringmann:2014lpa}. Although such a signal is not observed, our knowledge of the properties of the DM halo and the magnetic fields in the central region of the Galaxy is too limited to exclude the leptonic annihilation channel and, in the following, we will therefore exploit the possibility offered by the $\tau^+\tau^-$ final state.

In the extension of the type-X 2HDM that we consider, a pair of DM particles $S$ can annihilate into a pair of SM particles either via $H$ or the SM Higgs boson $h$. A further channel, yielding four SM particle final states, is provided by the $SS \to AA$ interaction, with the pseudo-scalar $A$ decaying promptly as $A \to \tau^+\tau^-$ and/or $A \to b \bar b$. 
As a DM particle lighter than the SM vector bosons provides a better fit of the GCE,\footnote{See Table I and Figure 3 in ref.~\cite{Calore:2014nla}.} in our analysis we restrict the DM mass of $m_S$ accordingly and forbid heavier SM final states kinematically. Furthermore, below we set $\lambda_{\rm SL} = 0$ to bar the $SS \to AA \to SM$ channel, resulting in non-trivial and boosted SM final states which require a new and dedicated analysis in the context of GCE. We postpone the study of this channel to a future work, presenting however some remarks on the consequences of $\lambda_{\rm SL} \neq 0$ in the last subsection.

\begin{figure}[h]
	\centering
\includegraphics[width=0.95\textwidth]{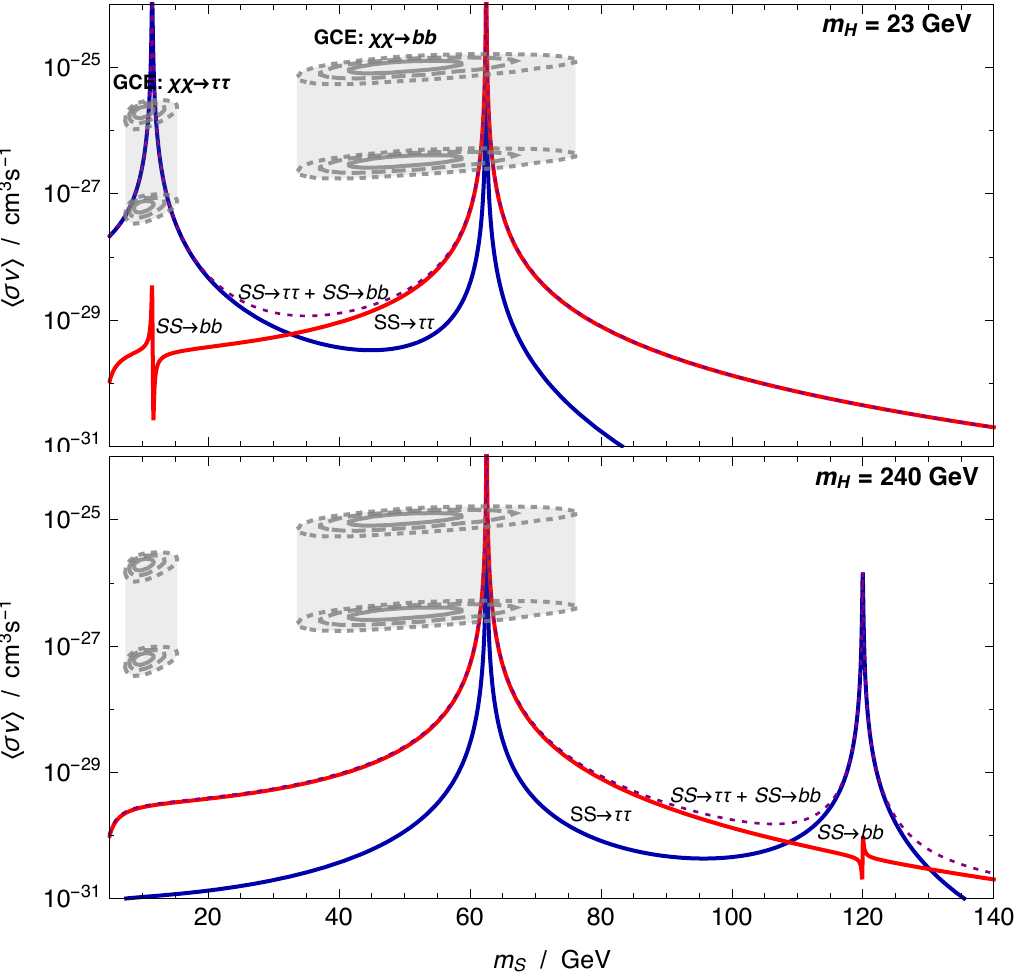}
\caption{The average annihilation cross section times the DM relative velocity for $SS \to \tau^+\tau^-$ (blue) and $SS \to b\bar{b}$ (red) versus the DM mass of $m_S$. The average is performed on the DM velocity distribution and the mass of $H$ is set to 23~GeV and 240~GeV in the upper and lower panel respectively. The remaining parameters are: $\tan \beta = 30$, $\lambda_{\rm SL} = 0$, $\lambda_{\rm SQ} = 0.004$, $m_A=20$~GeV and $\lambda_1=0.3$. The oval regions show the values of DM mass and $\sv$ that fit the GCE: $\tau^+\tau^-$ channel on the left and $b \bar b$ channel on the right. The grey regions show the systematic errors due to uncertainties in the DM halo profile~\cite{Calore:2014nla}. In the upper panel, our model fits both the GCE and the DM relic abundance through the $\tau^+\tau^-$ channel, satisfying at the same time the constraints from direct detection experiments and the SM Higgs invisible decay width. In the lower panel, through the $b \bar b$ channel, our model is able to fit the GCE as well as the DM relic abundance, satisfying also the constraints brought by direct detection experiments and by the SM Higgs invisible decay width. On top of that, in this case, our model  lowers as well the discrepancy that afflicts the prediction of the anomalous magnetic moment of the muon $g-2$ below the $\sim 2\sigma$.}
\label{fig:GCE1}
\end{figure}

Under the assumptions specified above, $S$ annihilates only via $H$ and SM Higgs boson $h$ into the dominant final states $b\bar b$ and $\tau^+\tau^-$. Figure~\ref{fig:GCE1} shows the quantitative behaviour of the corresponding cross sections $\sv$ multiplied by and averaged over the DM relative velocity.\footnote{The expressions for the annihilation and spin independent direct detection cross sections are given in~\cite{Okada:2013bna}.} The $b\bar b$ channel dominates for $m_S\sim m_h/2$ due to the hierarchy in the SM Yukawa couplings. The second peak, corresponding to $m_s \sim m_H/2$ is instead dominated by the $\tau^+\tau^-$ channel. In figure~\ref{fig:GCE1} we considered $\tb = 30$, and found that the effects of this parameter on the annihilation cross sections are small provided that $\tb>10$. We have set $\lambda_{\rm SL}=0$, as explained above, and $\lambda_{\rm SQ} = 0.004$. The constraints due to the SM Higgs boson invisible width and direct detection experiment require that $\lambda_{\rm SQ} < 0.006$. The parameters $\lambda_1$ and $m_A$ enter the mentioned annihilation cross sections only through the decay width of $h$ and $H$. We found that the values of the cross sections vary in a negligible way for $m_A \in [10, 250]$~GeV and $\lambda_1 \in [0, 2\pi/3]$.

\subsection{The $\tau^+\tau^-$ channel}

\begin{figure}[h]
\centering
\begin{subfigure}[b]{0.45\textwidth}
  \includegraphics[width=\textwidth]{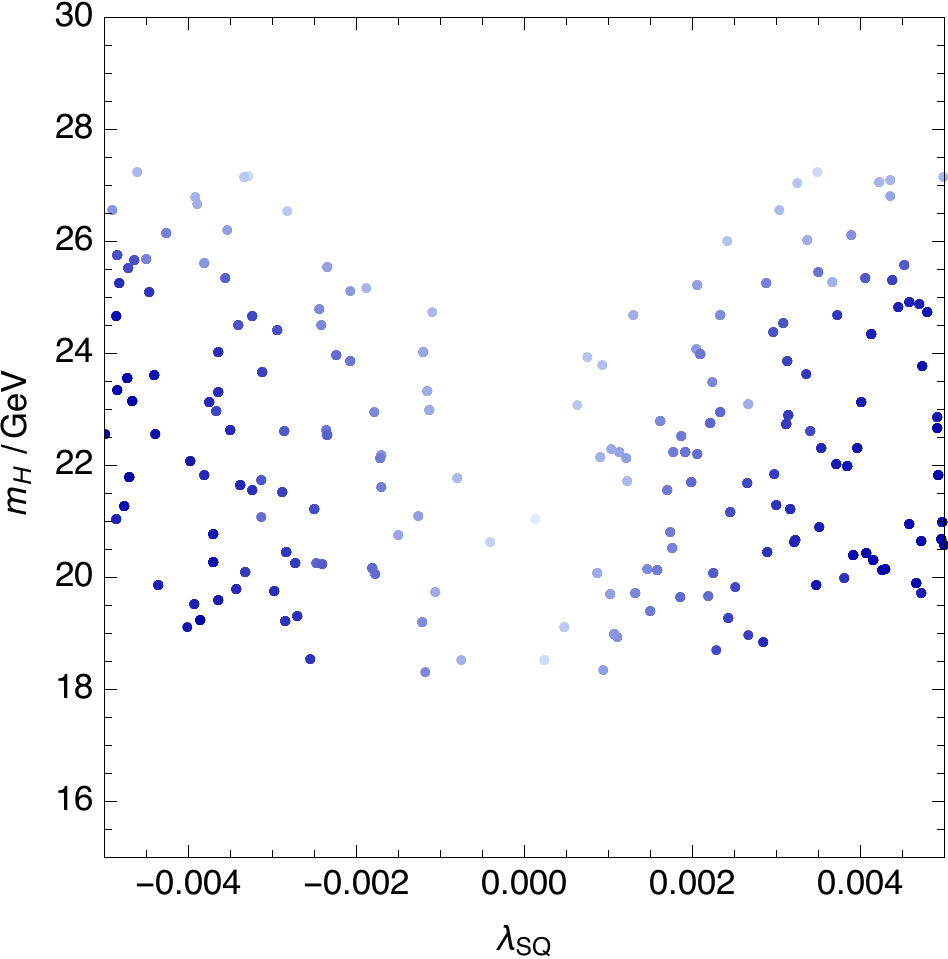}
  \caption{}
  \label{fig:GCE2a}
\end{subfigure}
~
\begin{subfigure}[b]{0.45\textwidth}
  \includegraphics[width=\textwidth]{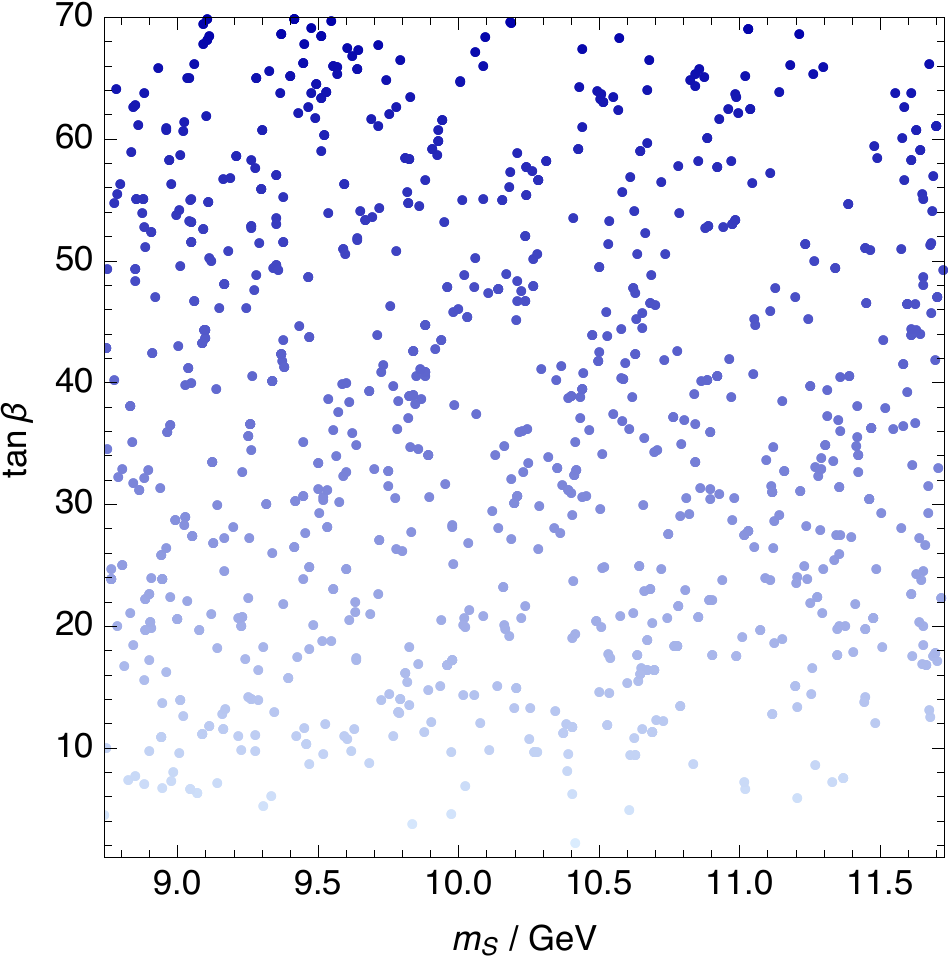}
  \caption{}
  \label{fig:GCE2b}
\end{subfigure}
\caption{Solutions of the model for the $\tau^+ \tau^-$ channel that fall within the 1$\sigma$ range (plus systematic uncertainties due to the DM halo profile) both of the annihilation cross section times velocity, required by the GCE fit, and of the DM relic abundance. The colour gradient denotes the value of $\tb \in[ 1,  70]$, with darker tones corresponding to higher values.}\label{fig:GCE2}
\end{figure}

\begin{figure}[h]
	\centering
\includegraphics[width=0.95\textwidth]{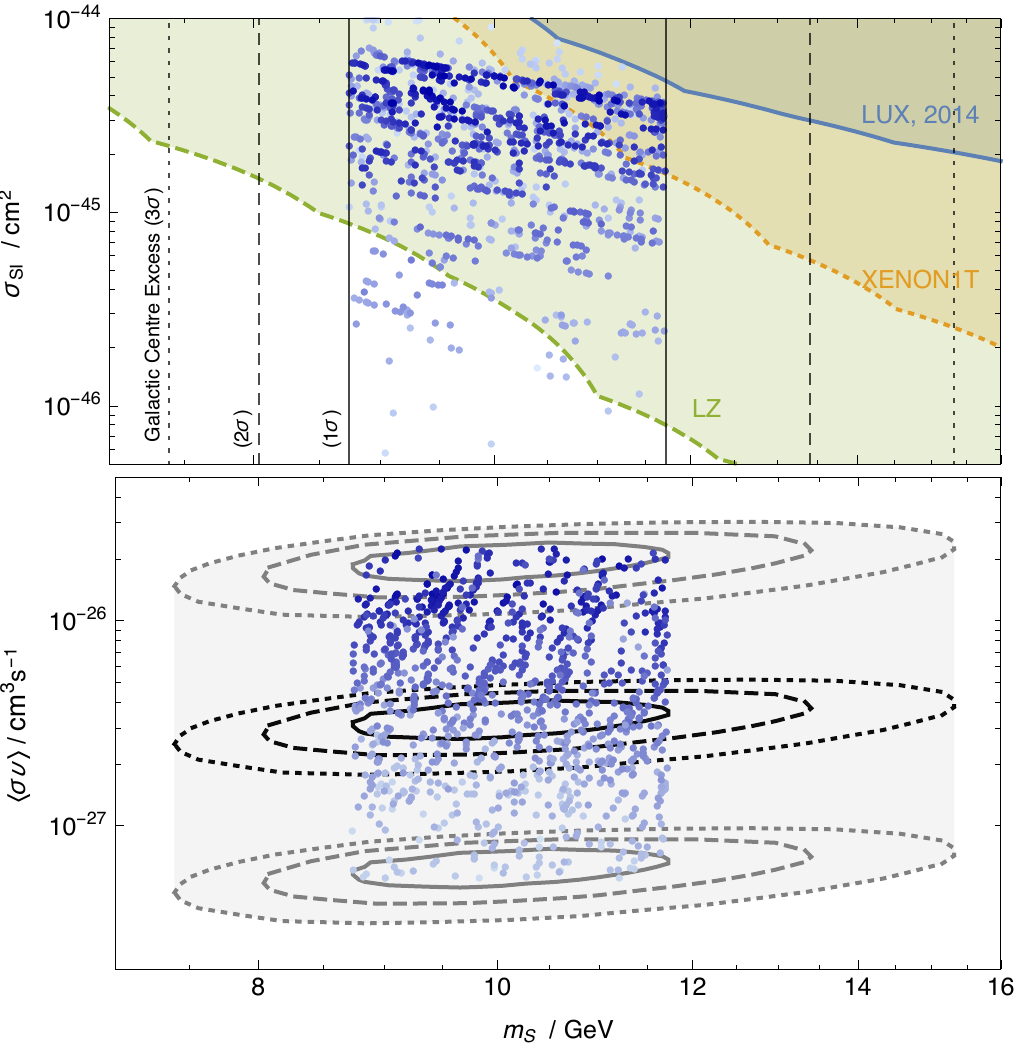}
\caption{The same as in figure~\ref{fig:GCE2b}, but projected on the planes  spanned by the DM mass $m_S$ and the spin independent direct detection cross section $\sigma_{\rm SI}$, upper panel, or the annihilation cross section times velocity $\sv$, lower panel. The coloured bands in the upper panel denote the constraints from the direct detection experiments LUX \cite{Akerib:2013tjd} , XENON1T \cite{2012arXiv1206.6288A} and LZ \cite{2011arXiv1110.0103M}. In the lower panel, the dark ellipses show the 1, 2 and 3$\sigma$ confidence level around the value required by the GCE fit. The grey region shows the uncertainty due to the limited knowledge of the DM halo profile~\cite{Calore:2014nla}.}\label{fig:GCE3}
\end{figure}

As indicated by the grey region on the left in figure~\ref{fig:GCE1}, fitting the GCE with the $\tau^+\tau^-$ channel requires a light DM particle $S$, $m_S = 9.96^{+1.05}_{-0.91}$~GeV, and an annihilation cross section times velocity $\sv_{\tau^+\tau^-} = \mathcal{A} \times 0.337^{+0.047}_{-0.048} \times 10^{-26}$~cm$^3$s$^{-1}$, where $\mathcal{A} \in [0.17 , 5.3]$ depending on the uncertainties in the DM halo profile at GC.\footnote{In this study we adopted the values in ref.~\cite{Calore:2014nla}}
Within the present framework such values can be achieved if the $S$ annihilation into $\tau^+\tau^-$ are mediated by $H$ and provided that the mass of the latter satisfies $m_H \approx 2 m_S$. Unfortunately, we find that such a light value of $m_H$ is in tension with the LEP bounds on light Higgs bosons \cite{Cao:2009as,Chang:2015goa} and does not allow for a combined fit of the GCE and of the anomalous magnetic moment of the muon, as discussed in section~\ref{sec:gm2}. Nevertheless, the $S$ annihilation into $\tau^+\tau^-$ provides a superb fit of the GCE as well as of the DM relic abundance. The methodology at the basis of our investigation is the following: we generated 400 random points within the parameter ranges $m_H \in [15,30]$~GeV and $\lambda_{\rm SQ}\in[-0.005, 0.005]$, in agreement with the constraint due to the invisible SM Higgs decay width and to the direct detection experiments. For every point $(m_H,\lambda_{\rm SQ})$ obtained in this way, we then generated other 2000 random points in the parameter space $m_S \in [8.74,  11.73]$~GeV, as required by the GCE at 1$\sigma$, and $\tb \in[ 1,  70]$. We finally calculated $\sv$ and the relic DM abundance for the resulting sample. Figure~\ref{fig:GCE2} shows where the points for which $\sv_{\tau^+\tau^-}$ and the relic DM abundance~\cite{Ade:2015xua} are within the corresponding 1$\sigma$ ranges fall on the $m_H$--$\lambda_{SQ}$ and $\tb$--$m_S$ planes. Figure~\ref{fig:GCE3} shows same points projected on the planes spanned by $m_S$ and either the spin-independent direct detection cross section $\sigma_{\rm SI}$ or the annihilation cross section times velocity $\sv_{\tau^+\tau^-}$.

\subsection{The $b\bar b$ channel}

\begin{figure}[h]
\centering
\begin{subfigure}[b]{0.45\textwidth}
  \includegraphics[width=\textwidth]{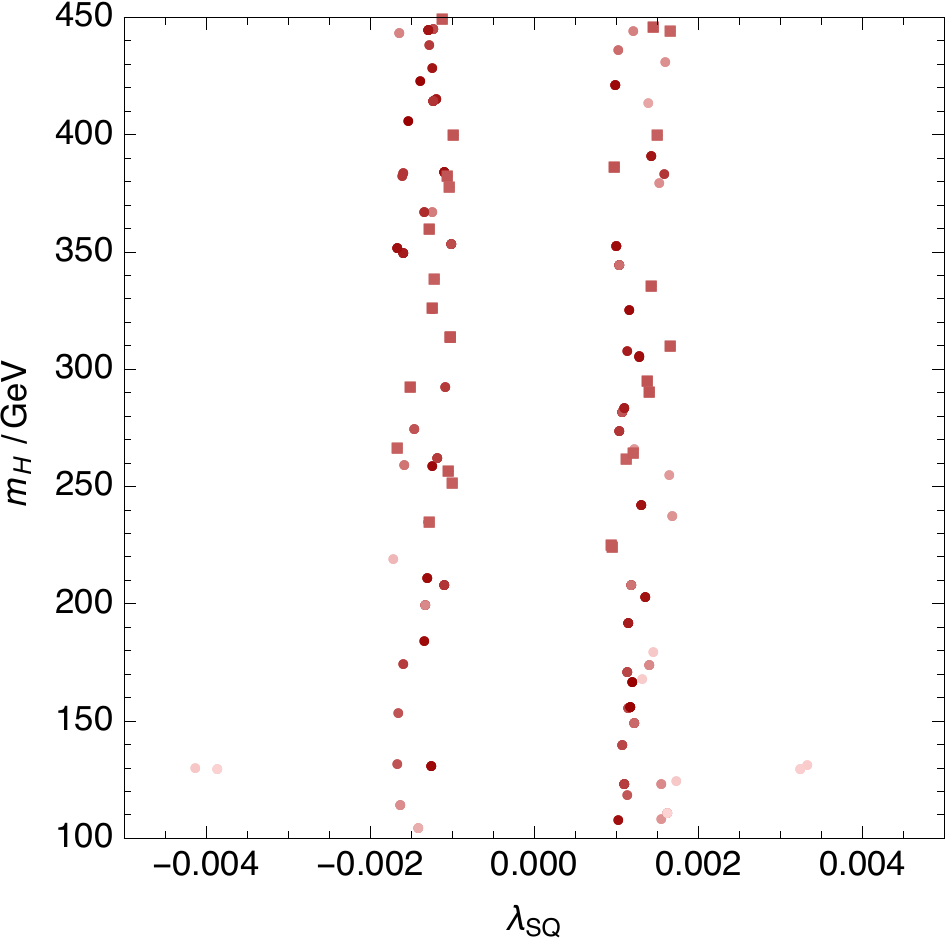}
  \caption{}
  \label{fig:GCE4a}
\end{subfigure}
~
\begin{subfigure}[b]{0.45\textwidth}
  \includegraphics[width=\textwidth]{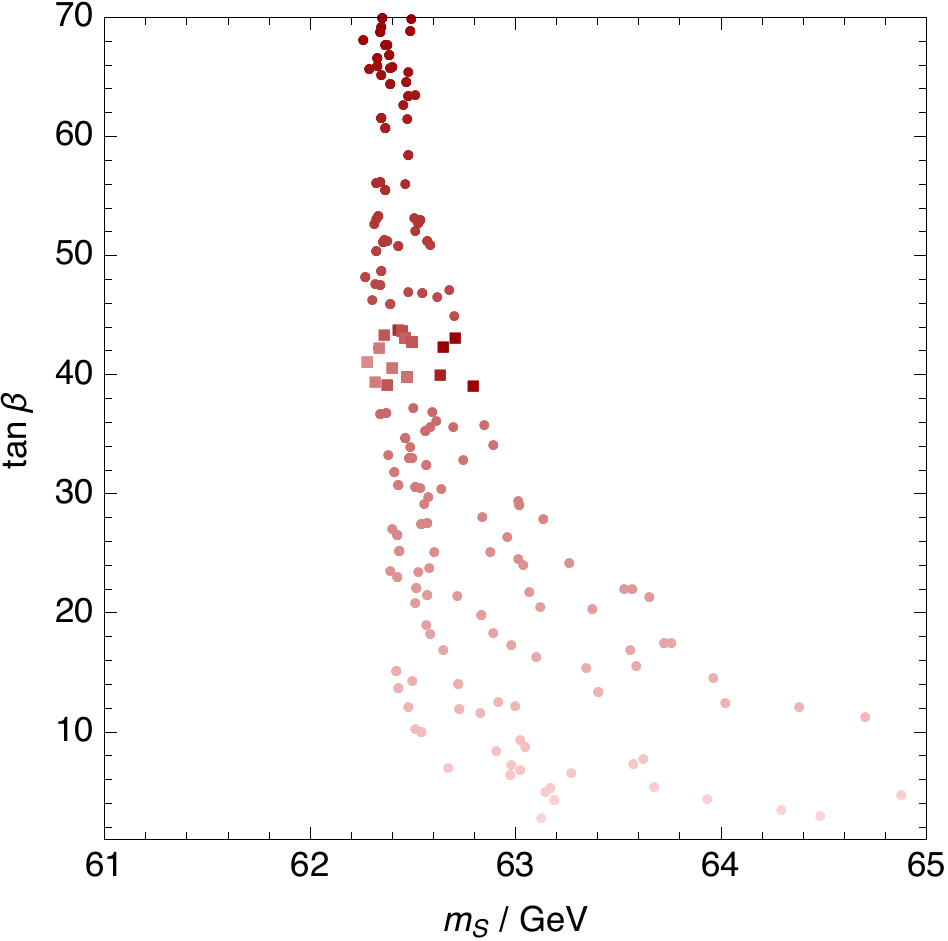}
  \caption{}
  \label{fig:GCE4b}
\end{subfigure}
\caption{Solutions of the model that fit the GCE at a 2$\sigma$ level (plus systematics due to DM halo profile uncertainties) and the DM relic abundance at 1$\sigma$ via the $SS \to b \bar b$ process. The colour gradient indicates the magnitude of $\tb \in[ 1,  70]$, with larger values in a darker tone. The rectangular points allow for the fit of the muon anomalous magnetic moment at the 2$\sigma$ confidence level.}\label{fig:GCE4}
\end{figure}

\begin{figure}[h]
	\centering
\includegraphics[width=0.95\textwidth]{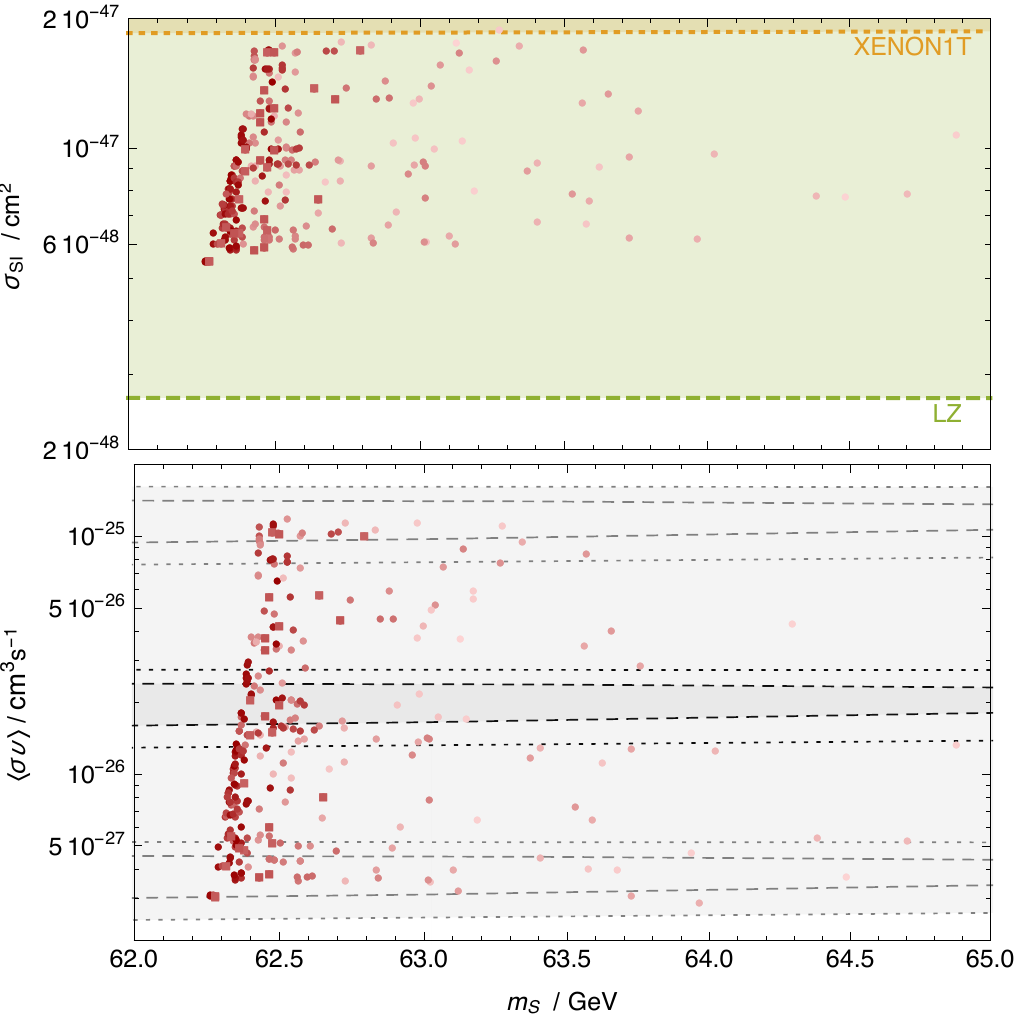}
\caption{The same points as in figure~\ref{fig:GCE4b} projected on the planes  spanned by the DM mass $m_S$ and the spin independent direct detection cross section $\sigma_{\rm SI}$, upper panel, or the annihilation cross section $\sv$, lower panel. The coloured bands in the upper panel denote the constraints from the indicated direct detection experiments. In the lower panel, the dark ellipses show the 2 and 3$\sigma$ confidence interval for the GCE fit while the grey region indicates the uncertainty due to the limit knowledge of the DM halo profile~\cite{Calore:2014nla}. The fit is always above the 1$\sigma$ level due to the condition $m_S \approx m_h/2$.}\label{fig:GCE5}
\end{figure}

Fitting the GCE with the $b \bar b$ channel yields a larger DM mass and annihilation cross section: $m_S = 48.7^{+6.4}_{-5.2}$~GeV and $\sv_{b \bar b} = \mathcal{A} \times 1.75^{+0.28}_{-0.26} \times 10^{-26}$~cm$^3$s$^{-1}$ with $\mathcal{A}$ defined above. Here $S$ annihilates mainly to $b \bar b$ via the SM Higgs $h$ and the required $\sv_{b \bar b}$ forces $m_S \approx m_h/2$, as shown in the lower panel of figure~\ref{fig:GCE1}. Adopting the $b \bar b$ channel to fit the GCE, the mass of $H$ can be large enough to allow also for the fit of muon anomalous magnetic moment, as shown in section~\ref{sec:gm2}. On the other hand, the quality of the GCE fit is poorer than in the $\tau^+\tau^-$ case, remaining at about the 1.5$\sigma$ level.
The result is obtaining by generating 800 random points on the parameter ranges $m_H\in[100,450]$~GeV, following the constraints of section~\ref{sec:gm2} and \ref{sec:Perturbative Unitarity and bounds from precision Electroweak measurements}, and $\lambda_{\rm SQ}\in[-0.005,0.005]$. For every point  $(m_H,\lambda_{\rm SQ})$ we then generated 2000 random points covering the parameter space $m_S\in[61,65]$~GeV, within the 2$\sigma$ range of the GCE best fit, and $\tb \in [1,70]$. For each member of the resulting sample we calculate $\sv$ and the relic DM abundance. Figure~\ref{fig:GCE4} shows the points yielding a $\sv_{b \bar b}$ within the corresponding 2$\sigma$ range that also reproduce the desired DM abundance~\cite{Ade:2015xua} within 1$\sigma$. The rectangular points denote the solutions for which our model's prediction of anomalous magnetic moment of the muon falls within the experimental 2$\sigma$ confidence level. Figure~\ref{fig:GCE5} shows same points projected on the plane spanned $m_S$ and the spin independent direct detection cross section $\sigma_{\rm SI}$ or the averaged annihilation cross section times velocity $\sv_{b \bar b}$.

\subsection{Allowing for both the $\tau^+\tau^-$ and $b \bar b$ channels}

The framework we consider allows in principle for DM annihilation proceeding through both the $\tau^+\tau^-$ and $b \bar b$ channels, with a branching that can be as large as $\mathcal{O}(1)$. We find however that having set $\lambda_{\rm SL} = 0$ in order to avoid the $SS \to AA$ annihilations, such a branching is never sizeable in our framework. The $\tau^+\tau^-$ and $b \bar b$ final states thus dominate the signal in complementary regions of the parameter space, as shown in figure~\ref{fig:GCE1}, justifying the presented analysis in which the subdominant channel is disregarded. 
Interestingly, relaxing the condition $\lambda_{\rm SL} = 0$ yields instead a sizeable branching between the considered channels and, therefore, a mixed $\tau^+\tau^-$ and $b \bar b$ final state that could be used to fit the GCE signal. Remarkably, as shown in figure~\ref{fig:GCE1}, this mixed final state has the potential to improve the quality of the GCE fit: whereas the low energy part of the latter is well described by DM annihilation into the $b \bar b$ final state, the complementary high energy power-law tail~\cite{Calore:2014nla, Calore:2014xka} is better fitted by the $\tau^+\tau^-$ channel. The quantitative analysis of the GCE in terms of mixed final states, as well as the study of the $SS\to AA \to SM$ channel, require the development of new dedicated tools that goes beyond the purpose of the present paper and is therefore postponed to a future work. 

\begin{figure}[h]
	\centering
\includegraphics[width=0.85\textwidth]{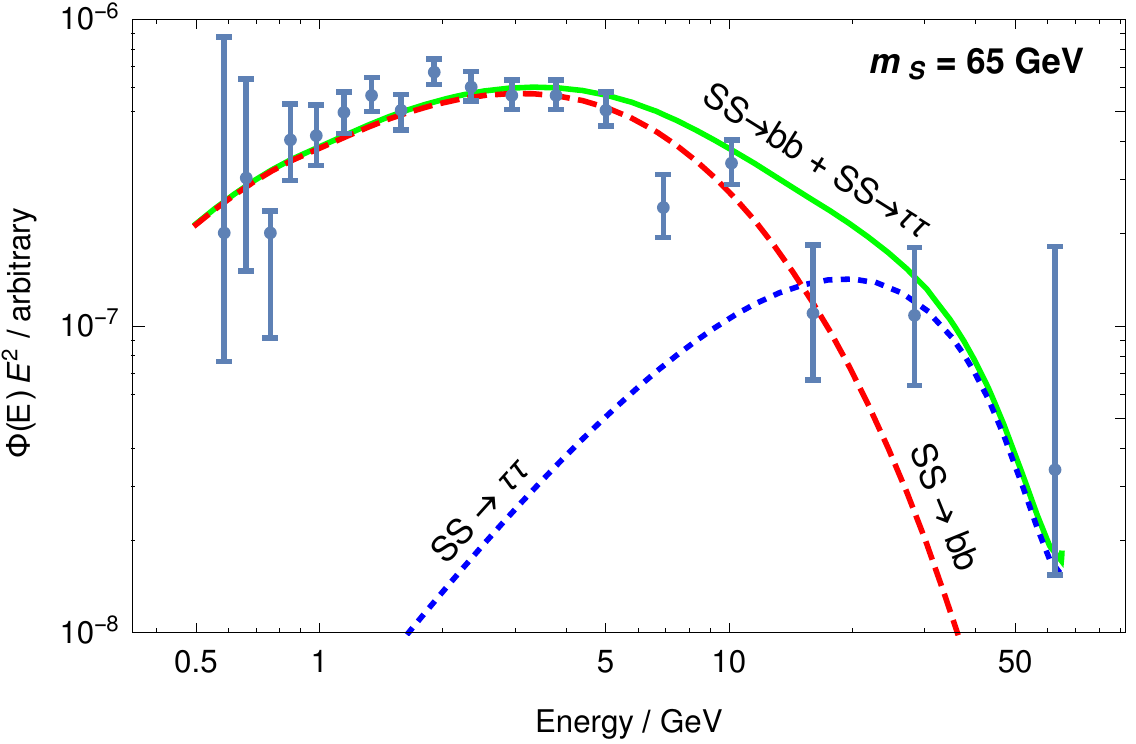}
\caption{Visual fitting of the prompt $\gamma$-ray spectrum from mixed annihilations to $b \bar b$ and $\tau^+\tau^-$ against the data of GCE. The extra annihilation to $\tau^+\tau^-$ seem to improve the quality of fit especially for the high energy tail of the signal. The branching ratio of the channels $b \bar b$ to $\tau^+\tau^-$ is fixed here to 0.3.}
\label{fig:GCE6}
\end{figure}


\section{Impact on the anomalous magnetic moment of the muon} 
\label{sec:gm2}
We focus now on the tension between the Standard Model prediction and the measurements of the muon anomalous magnetic moments $a_\mu := (g_\mu - 2) / 2$. Dedicated experiments set the latter to \cite{Bennett:2006fi,Beringer:1900zz}
\begin{equation}
	a_\mu^{\rm exp} = 116592091\,(54)(33) \times 10^{-11}
\end{equation}
while the corresponding Standard Model prediction, which comprises the QED, Electroweak and Hadronic corrections, matches \cite{Hagiwara:2011af,Broggio:2014mna,Davier:2010nc}
\begin{equation}
	a_\mu^{\rm SM} = 116591829\,(57) \times 10^{-11},
\end{equation}  
giving rise in this way to a discrepancy between experiments and theory which is more than $3\sigma$ large: 
\begin{equation}
	\label{eq:expdiscrgm2}
	\Delta a_\mu := a_\mu^{\rm exp} - a_\mu^{\rm SM} = 262 \, (85) \times 10^{-11}.
\end{equation}
 
Clearly, the physics beyond the Standard Model may help reducing this gap. Within the framework we propose, for instance, the anomalous magnetic moment of the muon receives additional contributions from both the extended Higgs and the dark sectors. Unfortunately, at the precision level achieved by current experiments, the impact of the latter is negligible given that the scalar field $S$ enters the computation of $\dmu$ starting at the two-loop level and that, differently from the case of the Barr-Zee diagrams \cite{Wu:2001vq, PhysRevLett.65.21} discussed below, the corresponding amplitudes do not benefit from any enhancement factor that would compensate the higher loop suppression. The calculation of the anomalous magnetic moment of the muon then proceeds here in the same fashion as in a pure lepto-specific 2HDM, with the first contribution to $a_\mu$ brought at the one-loop level by the extended Higgs sector \cite{Jegerlehner:2009ry}:
\begin{equation}
	\label{eq:damu1loop}
	\delta a_\mu^{(2HDM,1\ell)}
	=
	\frac{m^2_\mu}{8\pi^2 v^2}
	\!\!\!\!\!\!
	\sum_{j\in\{h, H, A, H^\pm\}}
	\!\!\!\!\!\!
	(y^j_\mu)^2 r^j_\mu F_j\left(r^j_\mu\right). 
\end{equation}
In the above formula $r^j_\mu:=m^2_\mu / M_j^2$, $y^j_\mu$ are the rescaled Yukawa couplings of the muon, presented in table~\ref{tab:2HDM:types}, while the loop functions $F_j$ are given by
\begin{align}
	F_{h, H}(y)
	&:=
	\int\limits_0^1\frac{x^2(2-x)}{1-x+yx^2}\, \td x
	\xrightarrow{y\ll1}
	-\ln(y) -\frac76 +\mathcal{O}(y),
	\\
	F_{A}(y)
	&:=
	\int\limits_0^1\frac{-x^3}{1-x+yx^2}\, \td x
	\xrightarrow{y\ll1}
	\ln(y) +\frac{11}{6} +\mathcal{O}(y),
	\\
	F_{H^\pm}(y)
	&:=
	\int\limits_0^1\frac{-x(1-x)}{1-(1-x)y}\, \td x
	\xrightarrow{y\ll1}
	-\frac16 +\mathcal{O}(y).
\end{align}
The one-loop correction brought by the extended Higgs sector is then dominated by the neutral scalars and pseudoscalar contributions, which interestingly have opposite signs. Notice that the boson $h$, which plays here the role of the Standard Model Higgs boson, is also included in the above summation. Our choice is motivated by the fact that whereas the pure Standard Model Higgs contribution accounted for in $a_\mu^{\rm SM}$ is negligible \cite{Jegerlehner:2009ry}, the condition in eq.~\eqref{eq:b-a_condition}, beside addressing the $h\to AA$ decay width, yields small deviations of $\sin(\beta-\alpha)$ from the Standard Model limit $\sin(\beta-\alpha) = 1$ that possibly result in a sizeable contribution to $a_\mu$.

\begin{figure}[h]
  \centering    
  \includegraphics[width=.7\textwidth]{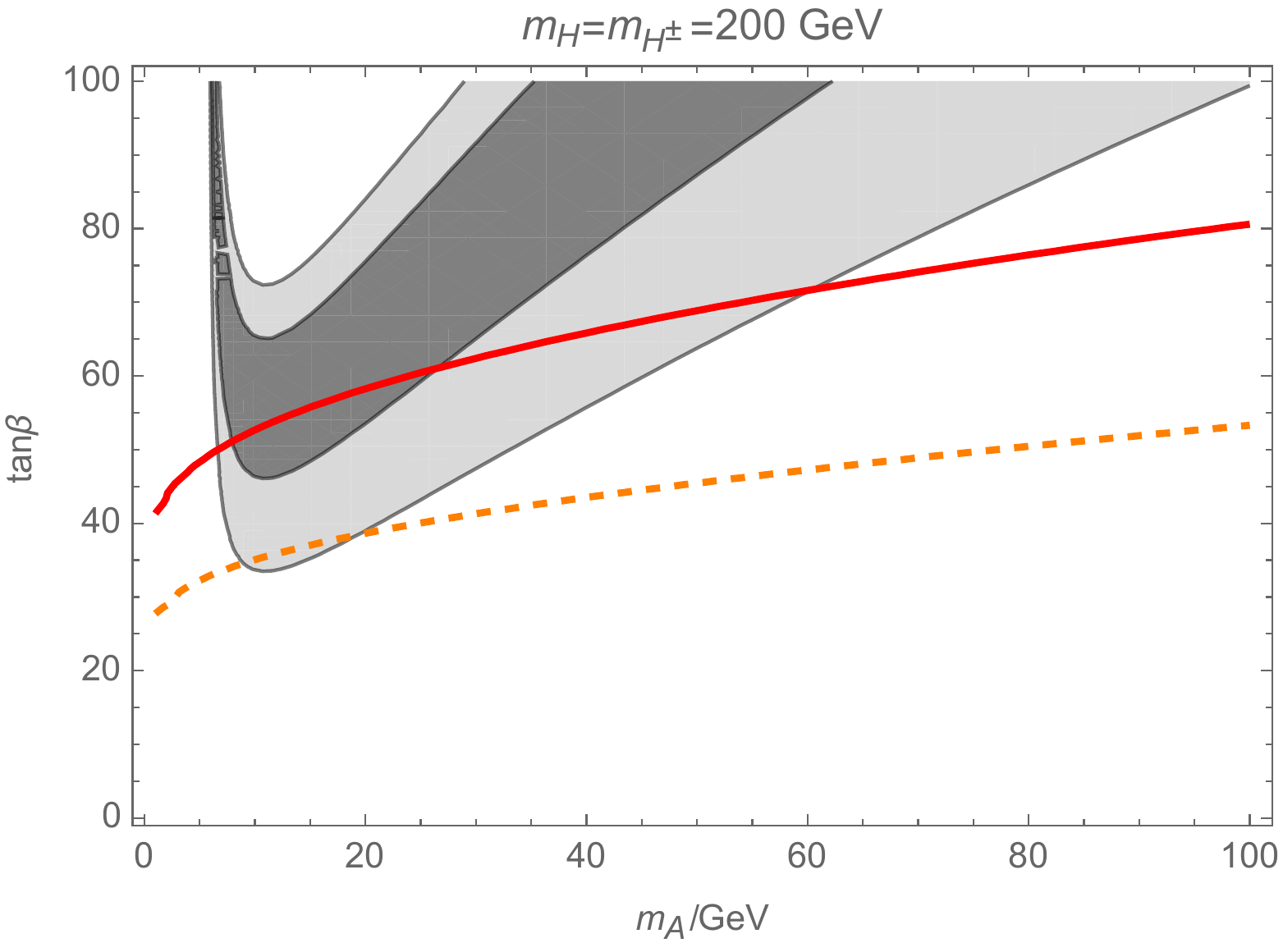}
  \\
  \includegraphics[width=.7\textwidth]{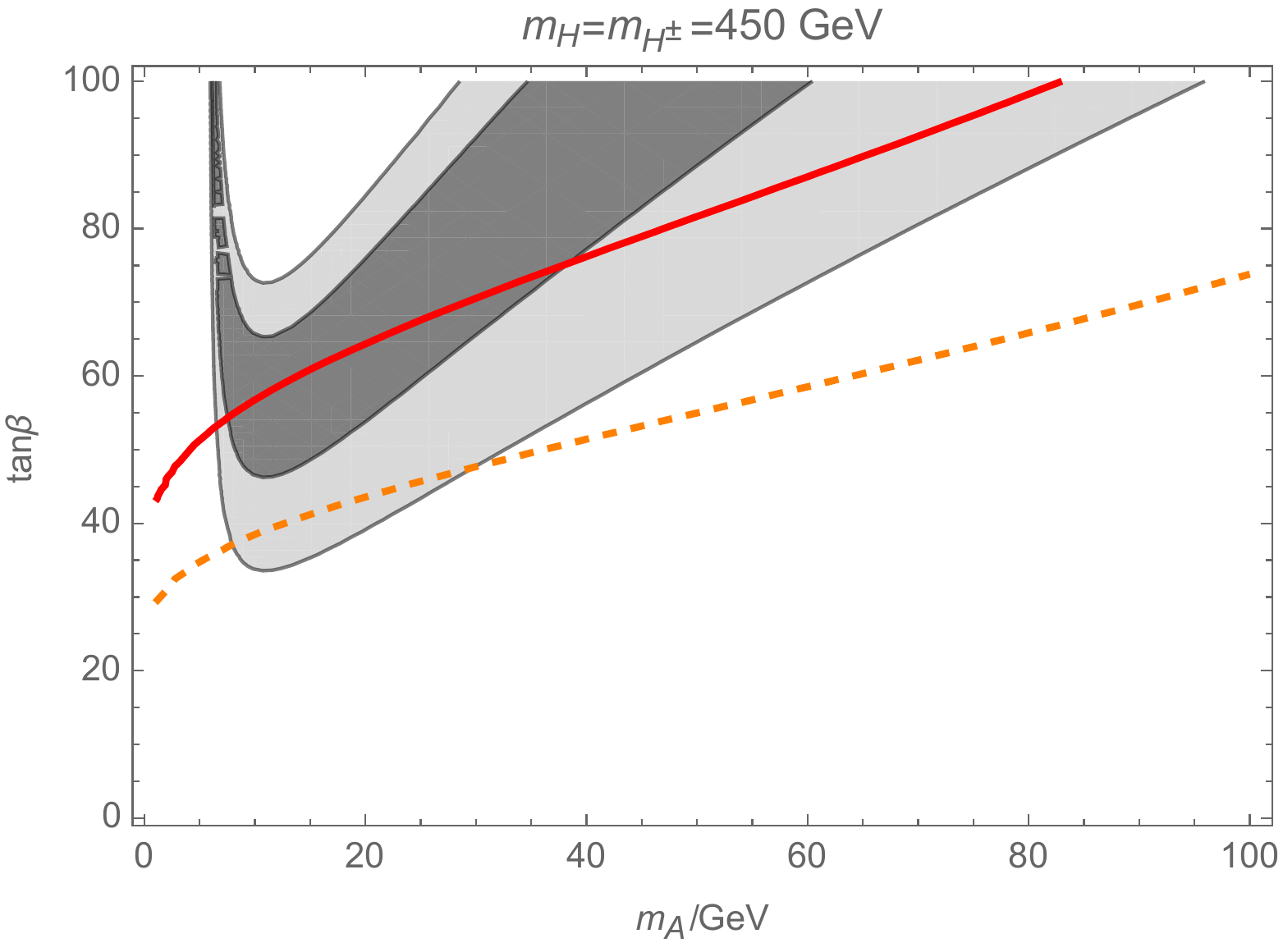}
  \caption{The contribution to the anomalous magnetic moment of the muon brought by the considered model as a function of $\tan\beta$ and the pseudoscalar mass $m_A$ for a fixed $m_H = m_{H^\pm}$. The areas shaded in dark and light grey correspond to the regions in which the discrepancy between theory and experiments is compatible with zero at the $68\%$ and $95\%$ confidence level respectively. The orange dashed line and the red line correspond to the $95\%$ and $99\%$ exclusion limits that lepton universality imposes \cite{Abe:2015oca}.}
  \label{fig:gm2}
\end{figure}

On top of the one-loop corrections, the anomalous magnetic moment of the muon is also particularly sensitive to the Barr-Zee diagrams that, at the two-loop level, induce an effective coupling between the neutral scalars or the pseudoscalar and the photons. These yield \cite{Chang:2000ii,Cheung:2003pw}  
\begin{equation}
	\label{eq:damu2loop}
	\delta a_\mu^{\rm(2HDM,\,BZ)}
	=
	\frac{\alpha m^2_\mu}{8\pi^3 v^2} 
	\!\!\!\!\!\!
	\sum_{\substack{i\in\{h, H, A, H^\pm\}\\ f\in\{t, b, \tau,\dots\}}}
	\!\!\!\!\!\!
	c_f \, q_f^2 \, y^i_\mu y^i_f r^i_f \, G_i\!\left(r^i_f\right)
\end{equation}
where $\alpha$ is the fine-structure constant, $r^i_f := m^2_f/M^2_i$, $m_f$ is the mass of a fermion $f$, while $c_f$, $q_f$ and $y^i_f$ are respectively its colour multiplicity, electric charge and rescaled Yukawa coupling. The loop functions are given by
\begin{align}
	G_{h,H}(y) & := \int\limits_0^1 \frac{2x(1-x)-1}{x(1-x)-y}\ln\left(\frac{x(1-x)}{y}\right) \td x,\\
	G_{A}(y) & := \int\limits_0^1 \frac{1}{x(1-x)-y}\ln\left(\frac{x(1-x)}{y}\right) \td x.	
\end{align}
The importance of the Barr-Zee diagrams stems from the $m^2_f/m^2_\mu$ enhancement of the contributions in eq.~\eqref{eq:damu2loop}, which can easily overcome the extra loop suppression factor that is absent in the corresponding one-loop contributions. Notice also that whereas the one-loop corrections brought by the (pseudo)scalars of the 2HDM are (negative) positive, the corresponding two loop corrections have opposite signs. Then, if the contributions of the 2HDM to $a_\mu$ are dominated by the scalar one-loop correction, it could be possible to bridge the discrepancy within the theoretical prediction of this quantity and the corresponding measurement. Alternatively, if the pseudoscalar corrections dominate and are determined by the (positive) Barr-Zee diagram, the model under consideration can also add on the SM prediction and effectively reduce the gap quantified in $\Delta a_\mu$. In order to investigate this possibility, we compare in figure~\ref{fig:gm2} the latter to the total contribution brought by the model: $\delta a_\mu^{\rm (2HDM)} = \delta a_\mu^{\rm(2HDM,1\ell)} + \delta a_\mu^{\rm(2HDM,\,BZ)}$.

In agreement with previous analyses \cite{Broggio:2014mna,Wang:2014sda,Abe:2015oca}, we find that within the considered framework a light pseudoscalar could in principle reconcile the theoretical prediction of $a_\mu$ with the corresponding measurement. As shown in figure~\ref{fig:gm2}, if the mass $m_A$ of this particle falls in the range $[10, 100]$ GeV and $\tan\beta \gtrsim 35$, the pseudoscalar contribution to $\delta a_\mu^{\rm (2HDM)}$ is large enough to significantly reduce the discrepancy $\Delta a_\mu$. The areas shaded in dark and light grey respectively signal the regions where the latter is compatible with a null discrepancy at the 68\% and 95\% confidence level. The proposed picture is due to the remaining particles in the extended Higgs sector being much heavier than the pseudoscalar, in a way that the corresponding corrections to the anomalous magnetic moment are suppressed. The result applies on the whole range considered for the parameter $m_H=m_H^\pm\in[200, 450]$ GeV. We also checked that the same values of the parameters result in negligible contributions to the anomalous magnetic moment of the electron, which remains well in agreement with its experimental measurement on the whole considered region of the parameter space.   

It was recently shown \cite{Abe:2015oca}, however, that lepton universality severely constrains the 2HDM pseudoscalar solution to the puzzle posed by the anomalous magnetic moment of the muon through the measurements of the leptonic $\tau$ decay widths. In the lepto-specific 2HDM, these quantities receive additional contributions from the $H^\pm$ decay channel, at the tree level, and from the remaining particles of the extended Higgs sector at higher orders. By computing the ratios of the $\tau$ leptonic decay widths evaluated at the one-loop level, it is then possible to constrain the 2HDM contribution through the experimental bounds on the lepton universality \cite{Amhis:2014hma}.
The procedure results in the 95\% and 99\% exclusion limit shown in figure~\ref{fig:gm2} by the orange-dashed and red line respectively, which bar a significant part of the parameter space associated to large $\tb$ values but weaken progressively for increasing values of $m_A$ and $m_H$.
We also remark that the current measurements of $B_s \to \mu\mu$ disfavour pseudoscalar with masses $m_A \lesssim 10$ GeV \cite{Logan:2000iv}. The constraint is relevant if the (heavy) neutral and charged Higgs masses are close to the lower limits of their ranges, while it is less severe for larger values of these parameters.  

The 2HDM pseudoscalar contribution can then reduce, but not completely remove, the discrepancy between prediction and measurement of the anomalous magnetic moment of the muon. As shown in figure~\ref{fig:gm2}, lowering the latter below the 95\% confidence limit requires a light pseudoscalar with mass $10$ GeV $\lesssim m_A \lesssim 30$ GeV and moderately large values of $\tan \beta$, $35\lesssim\tan\beta\lesssim 45$. On top of that, alleviating the constraints that lepton universality casts through the leptonic $\tau$ decay imposes a strongly split spectrum in the theory, with $m_A \ll 200 \text { GeV} \lesssim m_H = m_{H^\pm}$. Interestingly, this strict set of condition still remarkably allows for the fit of the GCE within the considered model. More in detail, the large value of $m_H$ indicated by the analyses of $a_\mu$ implies that the GCE signal results, in the present framework, from the process $SS \to b \bar b$. Our analyses revealed that the values of $\tan\beta$ necessary to reproduce the detected GCE signal, figure~\ref{fig:GCE4b}, are compatible with the requirement imposed by $a_\mu$.    

As for the possibility offered within the considered framework by a light neutral scalar $H$, by setting $m_A=m_H^\pm\in[100, 450]$ GeV we find that the constraints from lepton universality \cite{Abe:2015oca} forbid a substantial contribution to $\Delta a_\mu$ from a scalar as light as $m_H \approx 20$ GeV, as suggested by the GCE fit via the $\tau^+ \tau^-$ channel. As the corresponding exclusion exceeds the 3$\sigma$ level, we conclude that in this case it is not possible to address the problem of the anomalous magnetic moment of the muon. 


\section{Conclusions} 
\label{sec:Conclusions}

We considered the lepto-specific 2HDM augmented with a real singlet scalar $S$ that plays the role of DM. While the annihilations of $S$ into SM particles explain the GCE excess, the extended Higgs sector reduces the tension between prediction and measurement of the muon anomalous magnetic moment.

In this framework, the DM particles $S$ annihilate either via the neutral scalar $H$ or the SM Higgs boson $h$, mainly into $b\bar b$ and $\tau^+\tau^-$ respectively. These two possibilities correspond to different regions in the parameter space of the model: the $b\bar b$ channel requires $m_S\sim m_h/2$ whereas the $\tau^+\tau^-$ channel dominates for $m_S \sim m_H/2$. In order to fit the GCE with the mentioned channels, we scanned over the parameter regions given by $\tb \in [1,70]$, $m_A \in [10, 250]$~GeV, $\lambda_1 \in [0, 2\pi/3]$ and $\lambda_{\rm SQ}\in[-0.005, 0.005]$. For the $\tau^+\tau^-$ channel we took $m_H \in [15,30]$~GeV and $m_S \in [8.74, 11.73]$~GeV, while for the $b\bar b$ final state we considered $m_H\in[100,450]$~GeV and $m_S\in[61,65]$~GeV. The choice of these ranges is motivated by the constraints on the SM Higgs boson invisible width, by direct detection experiment and previous fits of GCE~\cite{Calore:2014nla}, as well as by the theoretical bounds due, for instance, to perturbative unitarity and vacuum stability. Beside analysing the GCE, we investigated whether it is possible to fit the muon anomalous magnetic moment in the considered regions of the parameter space. In the calculation of the latter we accounted for the one-loop and the two-loop Barr-Zee contributions brought by the additional neutral scalar $H$, charged scalars $H^\pm$ and pseudoscalar $A$ that appear in the considered model. 

With our method we find that:
\begin{itemize}
 \item The $\tau^+\tau^-$ channel fits the GCE and the DM relic abundance satisfying the constraints of direct detection and the Higgs invisible decay width. However, we find that because of the strong constraints imposed by lepton universality \cite{Abe:2015oca} through the measurements of the leptonic tau decays, it is not possible to address the puzzle of the anomalous magnetic moment of the muon in this setup.
 \item The $b\bar b$ channel fits the GCE, proposing a DM candidate which respects the bounds imposed by cosmology, direct detection experiments and Higgs invisible decays. Compared to the $\tau^+\tau^-$ channel, the $b\bar b$ final state result in a looser fit, remaining at the level of $\sim 1.5\sigma$ at best, mainly because of the restriction on the DM particle mass $m_S\sim m_h/2$ that the required annihilation cross section imposes.
On the other hand, the setup required to fit the GCE via the $b\bar b$ channel remarkably allows for sizeable contributions to the anomalous magnetic moment of the muon that survive the strict constraints imposed by lepton universality.
In particular, bringing the discrepancy between prediction and measurements of the latter within the 95\% confidence interval requires a light pseudoscalar with mass $10$ GeV $\lesssim m_A \lesssim 30$ GeV and moderately large values of $\tan \beta$, $35\lesssim\tan\beta\lesssim 45$.  
\end{itemize}
The next generation direct detection experiments, the LZ for example, have the potential to test both the proposed solutions. 

We remark that in our analyses we set $\lambda_{\rm SL} =0$ to bar the $SS \to AA \to \text{SM}$ channel, resulting in boosted SM final states that require a dedicated analysis which we delay to a future work. We however argued that relaxing this condition would result in a mixed $\tau^+\tau^-$ and $b \bar b$ final state that has potential to improve the quality of the GCE fit within the presented framework.


\appendix

\section{Renormalisation group equations}
\label{app:RGEs}

The beta functions for the gauge couplings, the top Yukawa coupling and scalar quartic couplings at one-loop level, defined by $d g_{i}/d \mu = \beta_{g_{i}}$, where $\mu$ is the renormalisation scale, are given by
\begin{equation}
\begin{split}
  16 \pi^{2} \beta_{g'} &= 7 g^{\prime 3},
  \\
  16 \pi^{2} \beta_{g} &= 7 g^{3},
  \\
  16 \pi^{2} \beta_{g_{3}} &= 7 g_{3}^{3},
  \\
  16 \pi^{2} \beta_{y_{t}} &= \left( -\frac{9}{4} g^{2} - \frac{17}{12} g^{\prime 2} - 8 g_{3}^2 \right) y_{t} + \frac{9}{2} y_{t}^3,
  \\
  16 \pi^{2} \beta_{\lambda_{1}} &= \frac{3}{8} \left(3 g^{4} + 2 g^{2} g^{\prime 2} + 3 g^{\prime 4} \right) - (9 g^{2} + 3 g^{\prime 2}) \lambda_{1} + 24 \lambda_{1}^2 + 2 \lambda_{3}^2 \\
  &+ 
 2 \lambda_{3} \lambda_{4} + \lambda_{4}^2 + \lambda_{5}^2 + 
 2 \lambda_{SL}^2,
  \\
  16 \pi^{2} \beta_{\lambda_{2}} &= \frac{3}{8} \left(3 g^{4} + 2 g^{2} g^{\prime 2} + 3 g^{\prime 4} \right) - (9 g^{2} + 3 g^{\prime 2}) \lambda_{2} + 24 \lambda_{2}^2 + 
 2 \lambda_{3}^2 \\
 &+ 2 \lambda_{3} \lambda_{4} + \lambda_{4}^2 + \lambda_{5}^2 + 12 y_{t}^2 \lambda_{2} +
 2 \lambda_{SQ}^2,
  \\
  16 \pi^{2} \beta_{\lambda_{3}} &= \frac{3}{4} (3 g^{4} - 2 g^{2} g^{\prime 2} + g^{\prime 4}) 
  - (9 g^{2} + 3 g^{\prime 2}) \lambda_{3} + 12 (\lambda_{1} + \lambda_{2}) \lambda_{3} + 4 \lambda_{3}^2 \\
  &+ 4 (\lambda_{1} + \lambda_{2}) \lambda_{4} + 2 \lambda_{4}^2 +  2 \lambda_{5}^2 + 6 y_{t}^2 \lambda_{3} + 4 \lambda_{SL} \lambda_{SQ},
  \\
  16 \pi^{2} \beta_{\lambda_{4}} &= 3 g^{2} g^{\prime 2} - 3 (3 g^{2} + g^{\prime 2}) \lambda_{4} + 4 ( \lambda_{1} + \lambda_{2}) \lambda_{4} + 8 \lambda_{3} \lambda_{4} + 4 \lambda_{4}^2 + 
 8 \lambda_{5}^2,
  \\
  16 \pi^{2} \beta_{\lambda_{5}} &= (-9 g^{2} - 3 g^{\prime 2} + 6 y_{t}^2 + 4 \lambda_{1} + 4 \lambda_{2} + 
   8 \lambda_{3} + 12 \lambda_{4}) \lambda_{5},
  \\
  16 \pi^{2} \beta_{\lambda_{S}} &= 72 \lambda_{S}^2 + 2 \lambda_{SL}^2 + 2 \lambda_{SQ}^2,
  \\
  16 \pi^{2} \beta_{\lambda_{SL}} &= -\frac{3}{2} (3 g^{2} + g^{\prime 2}) \lambda_{SL} + 12 (\lambda_{1} + 2 \lambda_{S}) \lambda_{SL} + (4 \lambda_{3} + 2 \lambda_{4}) \lambda_{SQ} 
  + 8 \lambda_{SL}^2,
  \\
  16 \pi^{2} \beta_{\lambda_{SQ}} &=-\frac{3}{2} (3 g^{2} + g^{\prime 2}) \lambda_{SQ}  + 12 (\lambda_{2} + 2 \lambda_{S}) \lambda_{SQ} + (4 \lambda_{3} + 2 \lambda_{4}) \lambda_{SQ} + 
 8 \lambda_{SQ}^2 \\
 &+ 6 y_{t}^2 \lambda_{SQ}.
\end{split}
\end{equation}

\acknowledgments
The authors thank A.~Fowlie, T.~Abe, R.~Sato and K.~Yagyu for useful discussions. This work was supported by grants PUT799, PUT808, IUT23-6, CERN+, TK120, MOBILITAS grants MJD387, MTT8, MTT60 due to the European Social Fund, and by the EU through the ERDF CoE program.

\bibliographystyle{JHEP}
\bibliography{2HDMS}

\end{document}